\documentclass{emulateapj}
\usepackage{natbib}
\usepackage{amsbsy}
\usepackage{apjfonts}

\newcommand{\tucgrugoodnobs}{$147$}
\newcommand{\tucqc}{$59$}
\newcommand{\gruqc}{$36$}

\newcommand{\tucfmem}{$0.14_{-0.04}^{+0.05}$}
\newcommand{\tucvmean}{$-129.1_{-3.5}^{+3.5}$}
\newcommand{\tucvdisp}{$8.6_{-2.7}^{+4.4}$}
\newcommand{\tucfehmean}{$-2.23_{-0.12}^{+0.18}$}

\newcommand{\tucfmemexpanded}{$0.14_{-0.04(-0.07)}^{+0.05(+0.1)}$}
\newcommand{\tucvmeanexpanded}{$-129.1_{-3.5(-8.1)}^{+3.5(+8.0)}$}
\newcommand{\tucvdispexpanded}{$8.6_{-2.7(-4.7)}^{+4.4(+12.1)}$}
\newcommand{\tucfehmeanexpanded}{$-2.23_{-0.12(-0.24)}^{+0.18(+0.53)}$}
\newcommand{\tucfehdispexpanded}{$0.23_{-0.13(-0.21)}^{+0.18(+0.48)}$}
\newcommand{\tucvmeantwoexpanded}{$37.1_{-10.5(-20.7)}^{+10.5(+20.6)}$}
\newcommand{\tucvdisptwoexpanded}{$75.9_{-7.0(-12.8)}^{+8.5(+17.5)}$}
\newcommand{\tucfehmeantwoexpanded}{$-0.66_{-0.08(-0.15)}^{+0.07(+0.15)}$}
\newcommand{\tucfehdisptwoexpanded}{$0.52_{-0.05(-0.09)}^{+0.06(+0.13)}$}
\newcommand{\tucvgradexpanded}{$0.7_{-0.5(-0.6)}^{+0.9(+2.5)}$}
\newcommand{\tucvthetaexpanded}{$-18.0_{-97.0(-150.0)}^{+96.0(+186.0)}$}
\newcommand{\tucfehgradexpanded}{$-0.01_{-0.02(-0.06)}^{+0.01(+0.01)}$}
\newcommand{\grufmem}{$0.12_{-0.04}^{+0.06}$}
\newcommand{\gruvmean}{$-140.5_{-1.6}^{+2.4}$}

\newcommand{\grufehmean}{$-1.42_{-0.42}^{+0.55}$}

\newcommand{\grufmemexpanded}{$0.12_{-0.04(-0.08)}^{+0.06(+0.12)}$}
\newcommand{\gruvmeanexpanded}{$-140.5_{-1.6(-7.2)}^{+2.4(+16.2)}$}
\newcommand{\gruvdispexpanded}{$2.9_{-2.1(-2.8)}^{+6.9(+73.0)}$}
\newcommand{\grufehmeanexpanded}{$-1.42_{-0.42(-0.79)}^{+0.55(+1.49)}$}
\newcommand{\grufehdispexpanded}{$0.41_{-0.23(-0.36)}^{+0.49(+1.19)}$}
\newcommand{\gruvmeantwoexpanded}{$19.7_{-14.2(-28.1)}^{+14.5(+28.5)}$}
\newcommand{\gruvdisptwoexpanded}{$85.3_{-9.6(-17.1)}^{+11.6(+24.7)}$}
\newcommand{\grufehmeantwoexpanded}{$-0.68_{-0.1(-0.2)}^{+0.1(+0.2)}$}
\newcommand{\grufehdisptwoexpanded}{$0.55_{-0.08(-0.14)}^{+0.09(+0.19)}$}
\newcommand{\gruvgradexpanded}{$1.2_{-0.7(-1.1)}^{+1.9(+6.2)}$}
\newcommand{\gruvthetaexpanded}{$-96.0_{-51.0(-77.0)}^{+224.0(+268.0)}$}
\newcommand{\grufehgradexpanded}{$-0.16_{-0.13(-0.37)}^{+0.09(+0.15)}$}
\newcommand{\gruvdispupperlim}{$9.8$}

\newcommand{\grufehdispupperlim}{$0.9$}
\newcommand{\tucfehdispupperlim}{$0.4$}
\newcommand{\gruvgradupperlim}{$3.1$}
\newcommand{\tucvgradupperlim}{$1.5$}
\newcommand{\grufehgradlowerlim}{$-0.3$}
\newcommand{\tucfehgradlowerlim}{$-0.0$}

\newcommand{\grumrhalfupperlim}{$2.5\times 10^6$}
\newcommand{\grumlratioupperlim}{$2645$}

\newcommand{\tucmrhalf}{$2.7_{-1.3}^{+3.1}\times 10^6$}
\newcommand{\tucmlratio}{$1913_{-950}^{+2234}$}

\newcommand{\tucvgrf}{$-201.5_{-2.3}^{+2.4}$}

\newcommand{\gruvgrf}{$-185.7_{-1.8}^{+2.1}$}

\newcommand{\tucmemtot}{$   8$}

\newcommand{\grumemtot}{$   7$}

\newcommand{\deriv}{\mathrm{d}}
\newcommand{\teff}{T_{\mathrm{eff}}}
\newcommand{\feh}{\mathrm{[Fe/H]}}
\newcommand{\alphafe}{[\alpha/\mathrm{Fe}]}
\newcommand{\logg}{\log g}

\newcommand{\los}{\mathrm{los}}

\usepackage{amsmath}

\begin{document}
\title{Magellan/M2FS Spectroscopy of Tucana 2 and Grus 1\footnote{This paper presents data gathered with the Magellan Telescopes at Las Campanas Observatory, Chile.}}
\shorttitle{Magellan/M2FS Spectroscopy of Tucana 2 and Grus 1}
\author{Matthew G. Walker\altaffilmark{1}, Mario Mateo\altaffilmark{2}, Edward W. Olszewski\altaffilmark{3}, Sergey Koposov\altaffilmark{4}, Vasily Belokurov\altaffilmark{4}, Prashin Jethwa\altaffilmark{4}, David L. Nidever\altaffilmark{2,5}, Vincent Bonnivard\altaffilmark{6}, John I. Bailey III\altaffilmark{2}, Eric F. Bell\altaffilmark{2}, Sarah R. Loebman\altaffilmark{2}}
\email{mgwalker@andrew.cmu.edu}
\altaffiltext{1}{McWilliams Center for Cosmology, Department of Physics, Carnegie Mellon University, 5000 Forbes Ave., Pittsburgh, PA 15213, United States}
\altaffiltext{2}{Department of Astronomy, University of Michigan, 311 West Hall, 1085 S. University Ave., Ann Arbor, MI 48109}
\altaffiltext{3}{Steward Observatory, The University of Arizona, 933 N. Cherry Ave., Tucson, AZ 85721}
\altaffiltext{4}{Institute of Astronomy, University of Cambridge, Madingley Road, Cambridge, CB3 0HA, United Kingdom}
\altaffiltext{5}{Large Synoptic Survey Telescope, 950 North Cerry Ave., Tucson, AZ 85719}
\altaffiltext{6}{LPSC, Universit\'e Grenoble-Alpes, CNRS/IN2P3, 53 avenue des Martyrs, F-38026, Grenoble, France}

\begin{abstract} 
We present results from spectroscopic observations with the Michigan/Magellan Fiber System (M2FS) of \tucgrugoodnobs\ stellar
targets along the line of sight to the newly-discovered `ultrafaint' stellar systems Tucana 2 (Tuc 2) and Grus 1 (Gru 1).  Based on simultaneous estimates of line-of-sight velocity and stellar-atmospheric parameters, we identify \tucmemtot\ and \grumemtot\ stars as probable members of Tuc 2 and and Gru 1, respectively.  Our sample for Tuc 2 is sufficient to resolve an internal velocity dispersion of \tucvdisp\ km s$^{-1}$ about a mean of \tucvmean\ km s$^{-1}$ (solar rest frame), and to estimate a mean metallicity of [Fe/H]= \tucfehmean.  These results place Tuc 2 on chemodynamical scaling relations followed by dwarf galaxies, suggesting a dominant dark matter component with dynamical mass \tucmrhalf\ $\mathrm{M}_{\odot}$ enclosed within the central $\sim 160$ pc, and dynamical mass-to-light ratio \tucmlratio\ $\mathrm{M}_{\odot}/L_{V,\odot}$.  For Gru 1 we estimate a mean velocity of \gruvmean\ km s$^{-1}$ and a mean metallicity of [Fe/H]=\grufehmean\, but our sample does not resolve Gru 1's velocity dispersion.  The radial coordinates of Tuc 2 and Gru 1 in Galactic phase space suggest that their orbits are among the most energetic within distance $\la 300$ kpc.  Moreover, their proximity to each other in this space arises naturally if both objects are trailing the Large Magellanic Cloud.
\end{abstract}


\section{Introduction}
\label{sec:intro}

A decade after the Sloan Digital Sky Survey revealed a population of `ultrafaint' Galactic satellites in the northern hemisphere \citep[e.g.,][]{willman05a,zucker06a,belokurov07}, a new generation of sky surveys has begun charting the southern contingent.  In the past year alone, nearly two dozen ultrafaint objects have been discovered using data from Pan-STARRS \citep{laevens15etal,laevens15betal}, the Dark Energy Survey \citep{koposov15,des15etal,des15betal,kim15} and other surveys using the Dark Energy Camera at Cerro Tololo \citep{kim15b,martin15etal}.  

\smallskip
The southern ultrafaints give new leverage to address old questions about the nature and origin of Galactic substructure.  While it continues to be revised, the faint end of the galactic luminosity function sets boundary conditions for galaxy formation within a given cosmological model \citep{koposov09}.  Complicating matters, the smallest (projected halflight radii $R_{\rm h}\la 30$ pc) and least luminous (absolute magnitude $M_{\rm V}\ga -4$) galaxies cannot be separated from globular clusters on the basis of luminous structural parameters alone.  Fortunately, stellar kinematics and chemical abundances reveal profound qualitative differences, as the smallest and least luminous dwarf galaxies tend also to have the largest dynamical mass-to-light ratios, while globular clusters typically do not exhibit evidence of dark matter \citep{mateo93,mateo98}.

\smallskip
Perhaps the most intriguing aspect of the newly-discovered objects is their apparent clustering near the Magellanic Clouds.  \citet[][`K15' hereafter]{koposov15} point out that the first batch of DES-detected objects lies, on average, closer to the Large Magellanic Cloud (LMC) than to the Milky Way.  Taking into account the spatial distribution of previously-known Galactic satellites, as well as selection effects due to uneven survey coverage, K15 go on to estimate a $\sim 6\%$ probability that the apparent clustering around the LMC arises randomly among unassociated objects.  From their data alone, \citet{des15betal} estimate a $< 0.1\%$ probability of observing such clustering if the satellite population is distributed isotropically around the Milky Way.  These apparent associations revive speculation regarding the Magellanic Clouds' viability as hosts for populations of their own dwarf-galactic satellites \citep{lynden-bell76,donghia08,deason15,yozin15}.  Given potentially profound implications for accounting of substructure within the Galactic halo, spectroscopic followup is required in order to classify the new objects and constrain their orbits.

\smallskip
Here we present an initial spectroscopic study of two ultrafaint systems discovered using data from the first year of the Dark Energy Survey: Tucana 2 and Grus 1.  Tuc 2 is detected as a stellar overdensity with $M_{\rm V}\sim -3.8$ and $R_{\rm h}\sim 165$ pc, at distance $\sim 57$ kpc (K15, \citealt[][`DES15' hereafter]{des15etal}).  Grus 1, originally spotted by K15 near a chip gap in public DES images, is less luminous ($M_{\rm V}\sim -3.4$), smaller ($R_{\rm h}\sim 62$ pc) and  farther away ($D\sim 120$ kpc).  The sizes derived from photometry alone suggest that both objects are dwarf galaxies, which typically have $R_{\rm h}\ga 30$ pc, rather than globular clusters, which typically have $R_{\rm h}\la 30$ pc \citep{gilmore07}.  We use our spectroscopic data to measure the systemic velocities and internal chemo-dynamical properties of these systems, as well as to identify individual member stars for followup investigations.

\section{Observations and Data Reduction}
\label{sec:obs}
\begin{figure}
  \includegraphics[width=4.in, trim=0.25in 2.5in 0.5in 0.5in,clip]{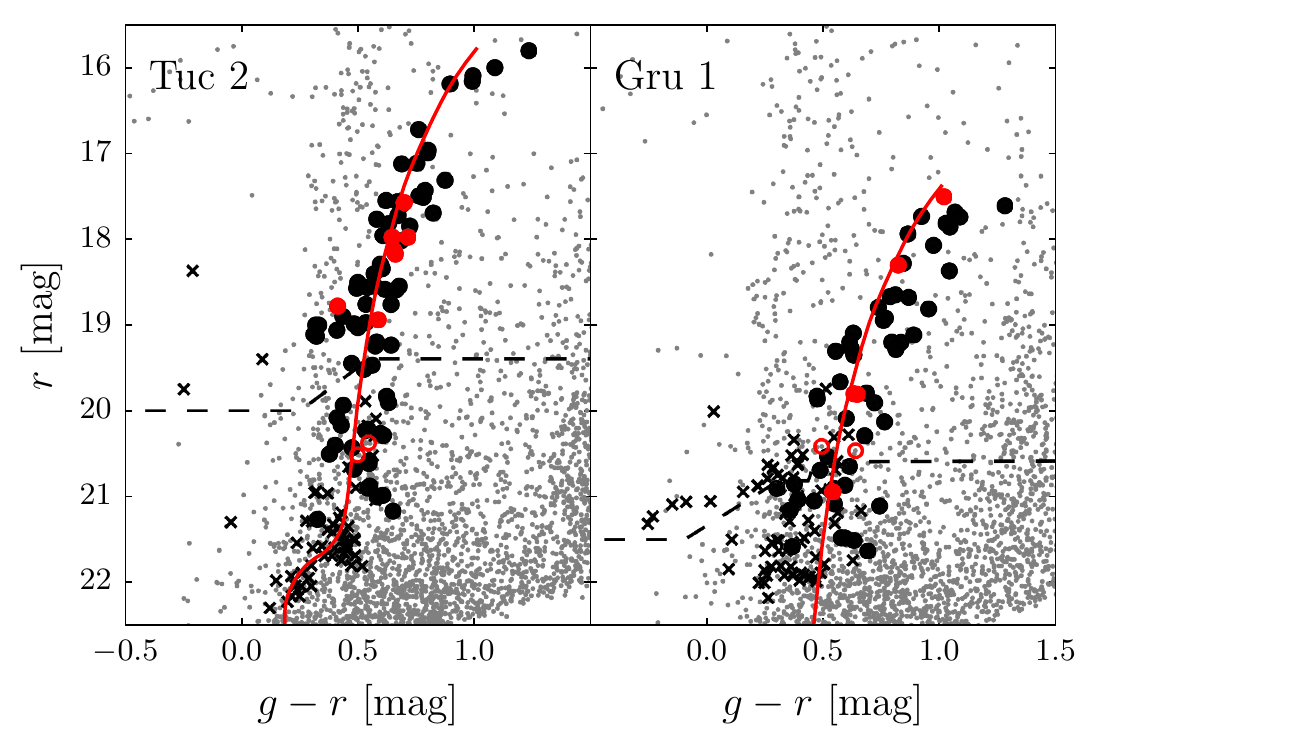}
  \caption{Color-magnitude diagrams for stars within $R\leq 15$ arcmin of the centers of Tuc 2 and Grus 1 \citep{koposov15}.  Red lines are Dartmouth isochrones \citep{dotter08} for age $=12$ Gyr, [Fe/H]$=-2.5$, [$\alpha/$Fe]=+0.4, and distance modulus $m-M=18.8$ (57 kpc) for Tuc 2 and $m-M=20.4$ (120 kpc) for Gru 1.  Large circles (`x's) indicate stars for which we measure velocities with errors $\leq 20$ km s$^{-1}$ ($>20$ km s$^{-1}$); filled circles are stars with observations passing additional quality-control criteria (Section \ref{subsec:membership}).  Dashed curves divide brighter targets observed with M2FS's high-resolution channel from the fainter ones observed with the medium-resolution channel.  Red/black colors identify probable members/nonmembers, based on positions and spectroscopy (see Section \ref{subsec:membership}).}
  \label{fig:tuc2gru1_cmd}
\end{figure}

We observed Tuc 2 and Gru 1 with the Michigan/Magellan Fiber System (M2FS; \citealt{mateo12}) at the 6.5-m Magellan/Clay telescope at Las Campanas Observatory, Chile, on the nights of 17 July (Tuc 2) and 18 July (Gru 1) 2015.  We obtained repeat observations of the Tuc 2 field on 12 September 2015.  We selected spectroscopic targets from the photometric catalogs that K15 generated from public images taken as part of the Dark Energy Survey (DES15), giving highest priority to point sources within 0.15 mag of metal-poor isochrones overlaid on color-magnitude diagrams (Figure \ref{fig:tuc2gru1_cmd}).  We also observed a handful of blue point sources in order to sample the horizontal branch.  

M2FS uses up to 256 fibers (entrance aperture of diameter $1.2$ arcsec) over a half-degree field, with fiber collision tolerance of 12 arcsec (center to center).  The fibers feed twin spectrographs that offer a variety of modes.  For both Tuc 2 and Grus 1, we observed the brightest targets (as well as horizontal branch candidates) in `HiRes' mode in one channel, covering the range $5132-5186$ \AA\ at effective resolution $\mathcal{R}\sim 18000$.  Simultaneously, we observed the faintest targets in `MedRes' mode in the other channel, deploying a new echellette grating to cover the range $5100 - 5315$ \AA\ at $\mathcal{R}\sim 10000$.  Immediately before and after science exposures, we acquired calibration spectra from a quartz lamp and Th-Ar arc lamp.  Also for the purpose of calibration, we acquired several twilight spectra at the beginning and end of each night.  

\begin{figure}
  \includegraphics[width=3.5in, trim=0.25in 2.5in 0.5in 0.5in,clip]{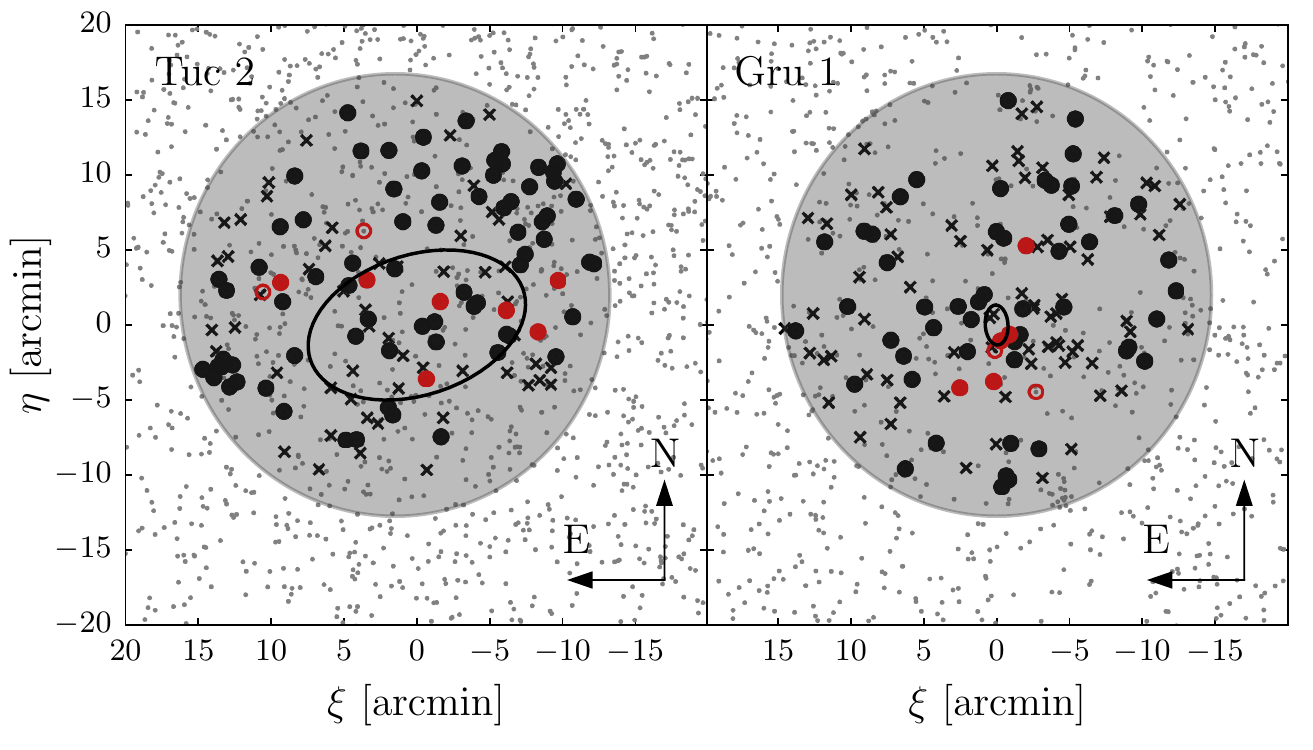}
  \caption{Standard coordinates for stars within $0.2$ magnitudes of the isochrones shown in the color-magnitude diagrams of Figure \ref{fig:tuc2gru1_cmd}.  Markers and their colors have the same meanings as in Figure \ref{fig:tuc2gru1_cmd}.  Black ellipses mark 2D elliptical halflight radii measured by \citet{koposov15}.  Large shaded circles represent the M2FS field of view.}
  \label{fig:tuc2gru1_map}
\end{figure}

\smallskip
Figure \ref{fig:tuc2gru1_cmd} displays color-magnitude diagrams for Tuc 2 and Gru 1, from the catalogs that K15.  Different symbols identify our spectroscopic targets, indicate quality of our spectroscopic measurements, and distinguish probable members from nonmembers.   Figure \ref{fig:tuc2gru1_map} shows the positions of these stars on the sky.  In the initial observing run (July 2015), we observed the Tuc 2 field for $5\times 1800$ s in below-average conditions (thin cirrus, median seeing $\sim 1.0$ arcsec) and the Gru1 1 field for $4\times 2400$ s in poorer conditions (similar seeing, patchy clouds).  In the followup run (September 2015), we observed the same Tuc 2 field for $5\times 2700$ s in variable but predominantly poor conditions (intermittent clouds, seeing $\ga 1.0$ arcsec).  

We processed all data frames using standard IRAF routines to extract one-dimensional, throughput-corrected spectra (see \citealt{walker15b} and Mateo et al., in preparation).  For HiRes and MedRes spectra, we used $\sim 30$ and $\sim 50$ emission lines, respectively, in the Th-Ar spectra to determine wavelength solutions that we then applied to the adjacent science exposures.  Residuals to wavelength solutions typically have rms $\sim 0.3$ km s$^{-1}$ (HiRes) and $\sim 0.7$ km s$^{-1}$ (MedRes).  Comparison of Th-Ar spectra acquired before and after science exposures indicate a drift of $\la 0.4$ km s$^{-1}$ (HiRes) and $\la 0.9$ km s$^{-1}$, implying that a temperature-dependent shift that affected previous M2FS observations of the Reticulum II dwarf galaxy \citep{simon15etal,walker15b} was negligible during observations of Tuc 2 and Gru 1 (ambient temperature inside the dome was more stable).  Finally, for each science frame we used the procedure of \citet{koposov11} to estimate the mean sky background ($\sim 30$ fibers in each channel were assigned to regions of blank sky), which we then subtracted from all science spectra.

\begin{figure}
  \includegraphics[width=4.in, trim=0.25in 0in 3.5in 3.in,clip]{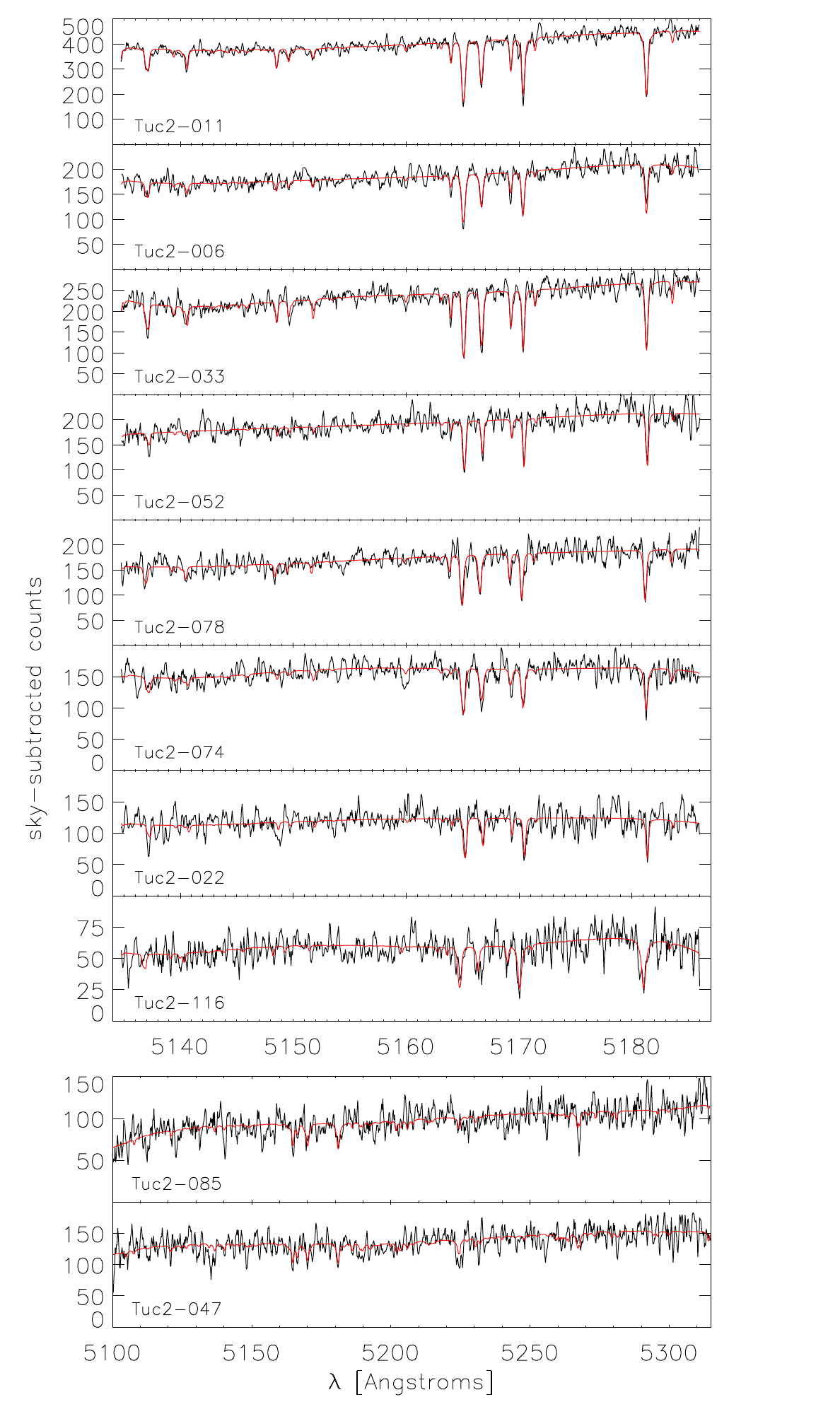}
  \caption{Sky-subtracted M2FS spectra (black) for probable members of Tuc 2, with best-fitting models overplotted (red).  The top eight spectra were acquired with the `HiRes' channel, the bottom two with `MedRes', during the July 2015 observing run.  Text gives the target ID (see Table \ref{tab:tuc2gru1_table1} for measured properties).}
  \label{fig:tuc2_specplot}
\end{figure}

\begin{figure}
  \includegraphics[width=4.in, trim=0.5in 0.7in 3.5in 3in,clip]{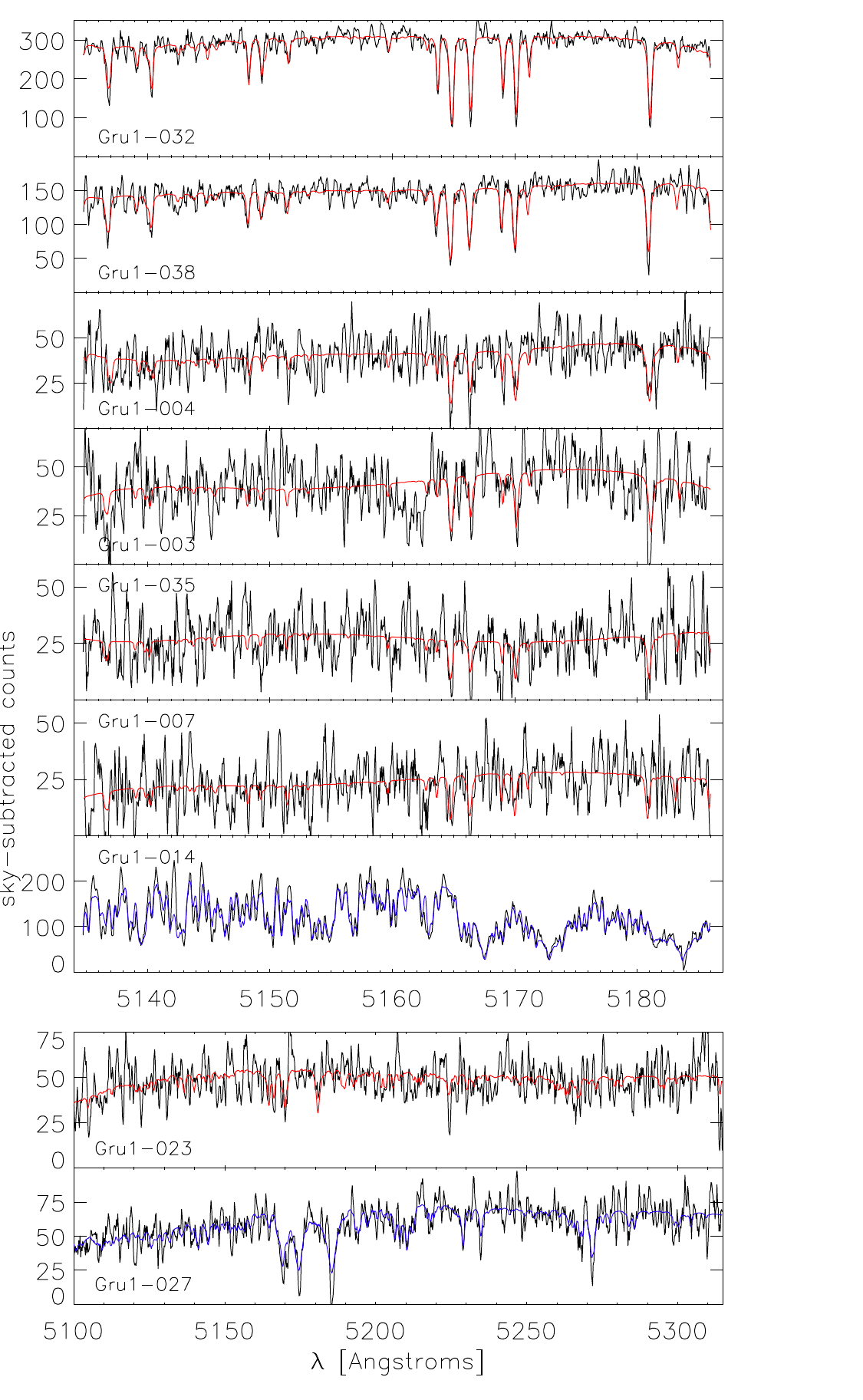}
  \caption{Same as Figure \ref{fig:tuc2_specplot}, but for probable members of Gru 1 (best-fitting models overplotted in red) and two examples of foreground interlopers (best-fitting models in blue).}
  \label{fig:gru1_specplot}
\end{figure}

\smallskip
Figure \ref{fig:tuc2_specplot} shows sky-subtracted M2FS spectra that we obtain from stacked science frames for ten probable members (see Section \ref{subsec:membership}) of Tuc 2.  Figure \ref{fig:gru1_specplot} shows spectra for seven probable members of Gru 1 and, for comparison, two probable interlopers contributed by the Galactic foreground.  Best-fitting model spectra are overplotted in each panel.

\section{Data}
\label{sec:data}

Following the Bayesian analysis described by \citet[][`W15' hereafter]{walker15,walker15b}, we use the MultiNest software package \citep{feroz08,feroz09} to fit each sky-subtracted spectrum with a model that is based on a library of synthetic spectra originally generated for the SEGUE Stellar Parameter Pipeline \citep[][`SSPP' hereafter]{lee08a,lee08b}.  The library spectra are calculated under the assumption that the abundance ratio of $\alpha$ elements to iron depends on iron abundance in the fllowing way: $\alphafe=+0.4$ for library spectra with $\feh < -1$, $\alphafe$ then decreases linearly as metallicity increases from $-1\leq \feh< 0$, and then $\alphafe=0$ for $\feh\geq 0$.  Our fitting procedure differs in one respect from that of W15: we adopt a log-normal rather than a uniform prior for effective temperature.  We determine the mean and variance of this prior, independently for each star, using the same relation between DES $g-r$ color and (logarithm of) SSPP effective temperature fit by \citet{koposov15betal}.  Given the narrow wavelength range of our spectra, posterior PDFs for $\teff$ tend to be dominated by the prior, thereby propagating valuable information derived from photometry.  

\smallskip
MultiNest returns random samplings from posterior probability distributions for line-of-sight velocity ($v_{\los}$), effective temperature ($\teff$), surface gravity ($\logg$), metallicity ($\feh$) and eleven additional nuisance parameters that let us simultaneously fit the continuum, adjust variance spectra, correct for systematic differences between wavelength solutions of target and template spectra, and broaden template spectra according to the instrumental resolution.  For all physical parameters, we record the first four moments of each posterior PDF: mean, variance, skewness and kurtosis.  We then adjust all means and variances according to results from our $\sim 500$ twilight spectra, which we use to estimate zero-point offsets (and uncertainties therein) with respect to known solar values (see W15 for details.  Mean offsets (standard deviations) for the high-resolution channel are $\langle\teff-\teff$$_{\odot}\rangle=-1$ K (0.001 K), $\langle\logg-\logg_{\odot}\rangle=-0.3$ dex (0.07 dex), $\langle\feh-\feh_{\odot}\rangle=-0.32$ dex (0.03 dex).  For the medium-resolution channel these offsets (standard deviations) are $\langle\teff-\teff$$_{\odot}\rangle=-1$ K (0.001 K), $\langle\logg-\logg_{\odot}\rangle=+0.05$ dex (0.03 dex), $\langle\feh-\feh_{\odot}\rangle=-0.16$ dex (0.03 dex).

\smallskip
For our spectroscopic observations of Tuc 2 and Gru 1, the first seven columns of Table \ref{tab:tuc2gru1_table1} list target ID, equatorial coordinates, $g$- and $r$-band magnitudes from K15's catalogs, heliocentric Julian dates of observation and median S/N per pixel in the sky-subtracted M2FS spectrum.  Columns 8-11 list estimates of $v_{\los}$, $\teff$, $\logg$, $\feh$; central values and error-bars correspond to the mean and standard deviation of the posterior PDF (adjusted according to our calibrations based on twilight spectra), while values listed in parentheses are skewness and kurtosis ($S=0$ and $K=3$ for a Gaussian distribution).  The final column indicates probability of membership in Tuc 2 or Gru 1, based on the analysis described in Section \ref{subsec:membership}.   We note that our repeat observations of several Tuc 2 targets yield parameter estimates that generally show good agreement, given the formal errors.  In subsequent analysis that uses stars with multiple independent measurements, we merge those measurements by taking inverse-variance-weighted means.  

The online version of this article includes all sky-subtracted M2FS spectra and corresponding variance spectra (in .fits format), obtained during this program.  Also provided are complete results from our spectroscopic modelling, including samples from 15-dimensional posterior PDFs for each star.  

\begin{table*}
\scriptsize
\caption{M2FS Stellar Spectroscopy of Tucana 2 and Gru 1$^{a}$}
\begin{tabular}{@{}ccccccccccccccccccccccc@{}}
\hline
\\
ID&$\alpha_{2000}$&$\delta_{2000}$&$g$&$r$&HJD$^{b}$&S/N$^{c}$&$v_{\rm los}$&$T_{\rm eff}$&$\log_{10}[g/$(cm/s$^{2}$)]&$\mathrm{[Fe/H]}$&$P_{\rm member}$\\

& [hh:mm:ss]&[$^{\circ}$:$\arcmin$:$\arcsec$]&[mag]&[mag]&[days]&&[km s$^{-1}$]$^{d}$&[K]&[dex]&[dex]\\
\hline
Tuc2-006&22:51:43.06&-58:32:33.7&$ 18.78$&$ 18.12$&$  7221.82$&$ 11.9^{\rm H}$&$-128.5\pm       0.7^{(   0.0,    3.3)}$&$5050\pm 149^{( 0.2, 3.0)}$&$1.92\pm0.38^{(-0.2, 3.1)}$&$-2.48\pm0.19^{( 0.1, 2.9)}$&$1.00\pm 0.00$\\
&&&&&$  7278.50$&$ 14.5^{\rm H}$&$-127.9\pm       0.8^{(   0.6,    4.4)}$&$4985\pm 141^{( 0.3, 3.1)}$&$1.96\pm0.50^{(-0.1, 2.6)}$&$-2.77\pm0.18^{( 0.3, 2.9)}$&&\\
Tuc2-011&22:51:50.28&-58:37:40.2&$ 18.27$&$ 17.57$&$  7221.82$&$ 18.9^{\rm H}$&$-127.6\pm       0.5^{(   0.1,    2.9)}$&$4779\pm 101^{( 0.4, 3.1)}$&$2.04\pm0.23^{(-0.1, 3.2)}$&$-2.42\pm0.14^{( 0.3, 2.9)}$&$0.99\pm 0.09$\\
&&&&&$  7278.50$&$ 17.8^{\rm H}$&$-127.0\pm       0.5^{(  -0.1,    3.0)}$&$4728\pm  93^{( 0.4, 3.2)}$&$1.30\pm0.28^{(-0.1, 2.9)}$&$-2.35\pm0.12^{( 0.3, 3.1)}$&&\\
Tuc2-022&22:52:21.38&-58:31:07.3&$ 19.52$&$ 18.93$&$  7221.82$&$  6.4^{\rm H}$&$-120.2\pm       2.0^{(  -0.2,    2.8)}$&$5322\pm 171^{( 0.1, 3.0)}$&$1.95\pm0.73^{(-0.0, 2.5)}$&$-2.42\pm0.26^{(-0.1, 2.9)}$&$1.00\pm 0.00$\\
Tuc2-033&22:51:08.32&-58:33:08.1&$ 18.68$&$ 17.97$&$  7221.82$&$ 11.6^{\rm H}$&$-126.4\pm       0.5^{(  -0.3,    3.1)}$&$4814\pm 118^{( 0.3, 3.0)}$&$1.26\pm0.29^{(-0.3, 3.1)}$&$-2.21\pm0.16^{( 0.2, 2.9)}$&$0.99\pm 0.05$\\
&&&&&$  7278.50$&$ 12.4^{\rm H}$&$-128.5\pm       0.5^{(  -0.2,    3.0)}$&$4538\pm 109^{( 0.2, 3.0)}$&$1.28\pm0.29^{(-0.1, 3.0)}$&$-2.56\pm0.15^{( 0.3, 3.0)}$&&\\
Tuc2-047&22:52:22.99&-58:27:51.0&$ 20.91$&$ 20.37$&$  7221.82$&$  6.8^{\rm M}$&$-140.1\pm       7.2^{(  -0.5,    4.5)}$&$5561\pm 204^{( 0.2, 3.0)}$&$1.79\pm0.91^{( 0.2, 2.5)}$&$-1.60\pm0.32^{(-0.2, 3.5)}$&$0.50\pm 0.29$\\
Tuc2-052&22:50:51.63&-58:34:32.5&$ 18.83$&$ 18.17$&$  7221.82$&$  9.8^{\rm H}$&$-122.8\pm       0.7^{(  -0.1,    3.5)}$&$5206\pm 166^{( 0.1, 3.0)}$&$1.95\pm0.50^{(-0.3, 3.0)}$&$-2.61\pm0.21^{( 0.0, 2.8)}$&$1.00\pm 0.00$\\
&&&&&$  7278.50$&$ 11.6^{\rm H}$&$-123.9\pm       0.9^{(   0.7,    4.2)}$&$4966\pm 149^{( 0.3, 3.1)}$&$1.86\pm0.47^{(-0.1, 2.8)}$&$-2.71\pm0.19^{( 0.2, 3.0)}$&&\\
Tuc2-074&22:53:06.67&-58:31:16.0&$ 19.19$&$ 18.77$&$  7221.82$&$  9.4^{\rm H}$&$-126.0\pm       1.5^{(   0.0,    3.4)}$&$5877\pm 210^{( 0.1, 2.9)}$&$2.23\pm0.68^{(-0.2, 2.7)}$&$-1.87\pm0.25^{(-0.1, 2.9)}$&$0.42\pm 0.38$\\
&&&&&$  7278.50$&$  7.6^{\rm H}$&$-130.3\pm       1.8^{(  -0.0,    2.9)}$&$5887\pm 211^{( 0.2, 3.0)}$&$2.00\pm0.79^{( 0.1, 2.4)}$&$-1.75\pm0.27^{(-0.1, 2.9)}$&&\\
Tuc2-078&22:50:41.07&-58:31:08.3&$ 18.62$&$ 17.97$&$  7221.82$&$  8.9^{\rm H}$&$-135.0\pm       0.9^{(   0.1,    3.1)}$&$5090\pm 149^{( 0.3, 3.0)}$&$2.11\pm0.41^{(-0.3, 3.2)}$&$-2.14\pm0.19^{( 0.1, 2.9)}$&$0.89\pm 0.24$\\
&&&&&$  7278.50$&$ 11.3^{\rm H}$&$-135.8\pm       0.8^{(  -0.2,    3.4)}$&$4936\pm 133^{( 0.3, 3.1)}$&$1.96\pm0.35^{(-0.2, 3.1)}$&$-2.22\pm0.17^{( 0.2, 2.9)}$&&\\
Tuc2-085&22:53:15.90&-58:31:52.6&$ 21.01$&$ 20.51$&$  7221.82$&$  5.6^{\rm M}$&$-130.8\pm       8.9^{(  -0.0,    2.1)}$&$5532\pm 194^{( 0.1, 3.0)}$&$3.93\pm0.69^{(-1.0, 4.4)}$&$-2.00\pm0.39^{(-0.0, 2.9)}$&$0.94\pm 0.07$\\
Tuc2-116&22:53:34.11&-58:38:10.3&$ 19.87$&$ 19.23$&$  7221.82$&$  4.8^{\rm H}$&$-146.4\pm       2.5^{(  -0.0,    3.6)}$&$5268\pm 170^{( 0.1, 3.0)}$&$3.40\pm0.66^{(-0.6, 4.6)}$&$-2.02\pm0.31^{(-0.2, 3.1)}$&$0.75\pm 0.27$\\
Gru1-003&22:56:37.05&-50:10:24.8&$ 20.45$&$ 19.80$&$  7222.80$&$  2.7^{\rm H}$&$-137.6\pm      3.9^{(  -0.1,    2.4)}$&$5219\pm 179^{( 0.1, 2.9)}$&$3.26\pm1.08^{(-0.5, 2.8)}$&$-2.03\pm0.43^{( 0.1, 2.9)}$&$0.99\pm 0.03$\\
Gru1-004&22:56:40.78&-50:10:51.4&$ 20.43$&$ 19.80$&$  7222.80$&$  2.9^{\rm H}$&$-139.4\pm      1.4^{(   0.1,    3.5)}$&$5225\pm 179^{( 0.2, 3.0)}$&$2.18\pm0.81^{(-0.3, 2.5)}$&$-1.32\pm0.34^{( 0.0, 3.0)}$&$0.99\pm 0.04$\\
Gru1-007&22:56:43.20&-50:11:30.0&$ 21.10$&$ 20.46$&$  7222.80$&$  1.7^{\rm H}$&$-142.7\pm     12.3^{(  28.9, 1462.6)}$&$5240\pm 175^{( 0.1, 3.0)}$&$1.21\pm0.64^{( 0.9, 3.7)}$&$-1.19\pm0.42^{(-0.7, 6.7)}$&$0.97\pm 0.05$\\
Gru1-014&22:56:54.84&-50:11:32.6&$ 18.91$&$ 17.86$&$  7222.80$&$  7.6^{\rm H}$&$  19.5\pm      0.4^{(  -0.2,    3.2)}$&$4288\pm  69^{(-0.2, 2.9)}$&$4.68\pm0.33^{(-0.4, 2.9)}$&$-0.56\pm0.16^{(-0.0, 2.8)}$&$0.01\pm 0.04$\\
Gru1-023&22:56:43.79&-50:13:32.7&$ 21.48$&$ 20.94$&$  7222.80$&$  3.2^{\rm M}$&$-147.5\pm      3.3^{(   0.1,    4.2)}$&$5503\pm 190^{( 0.1, 3.0)}$&$2.13\pm1.05^{(-0.2, 2.1)}$&$-1.12\pm0.45^{( 0.2, 3.0)}$&$0.83\pm 0.17$\\
Gru1-027&22:57:09.40&-50:09:56.9&$ 21.26$&$ 20.65$&$  7222.80$&$  4.1^{\rm M}$&$ 119.8\pm      2.2^{(  -0.4,    3.0)}$&$5073\pm 102^{( 0.1, 2.9)}$&$4.73\pm0.22^{(-1.0, 3.9)}$&$-0.74\pm0.26^{(-0.0, 2.7)}$&$0.00\pm 0.01$\\
Gru1-032&22:56:58.06&-50:13:57.9&$ 18.52$&$ 17.51$&$  7222.80$&$ 13.0^{\rm H}$&$-138.4\pm      0.4^{(  -0.1,    3.1)}$&$4270\pm  69^{( 0.0, 3.0)}$&$0.72\pm0.22^{(-0.1, 2.4)}$&$-2.37\pm0.10^{( 0.2, 2.9)}$&$0.73\pm 0.40$\\
Gru1-035&22:56:25.70&-50:14:14.2&$ 20.90$&$ 20.41$&$  7222.80$&$  1.9^{\rm H}$&$-143.1\pm      5.3^{(  30.5, 2432.1)}$&$5587\pm 204^{( 0.1, 3.0)}$&$2.28\pm1.07^{( 0.1, 2.3)}$&$-1.08\pm0.42^{(-0.3, 3.8)}$&$0.73\pm 0.22$\\
Gru1-038&22:56:29.92&-50:04:33.3&$ 19.12$&$ 18.30$&$  7222.80$&$  9.1^{\rm H}$&$-144.3\pm      0.8^{(  -0.3,    3.1)}$&$4532\pm 100^{( 0.3, 3.2)}$&$0.87\pm0.31^{( 0.2, 2.6)}$&$-2.10\pm0.15^{( 0.2, 3.0)}$&$0.69\pm 0.37$\\
\hline
\end{tabular}
\\
\raggedright
$^{a}$This version lists results only for stars with spectra shown in Figures \ref{fig:tuc2_specplot} and \ref{fig:gru1_specplot}.  See electronic edition for complete data table.\\
$^{b}$ heliocentric Julian date minus $2.45\times 10^6$ days\\
$^{c}$median signal-to-noise ratio per pixel; superscript specifies HiRes (H) or MedRes (M) channel\\
$^{d}$line-of-sight velocity in the heliocentric rest frame\\
\label{tab:tuc2gru1_table1}
\end{table*}

\section{Results}
\label{sec:results}

\begin{figure*}
  \includegraphics[width=6.5in,trim=0in 0.3in 0in 1.75in,clip]{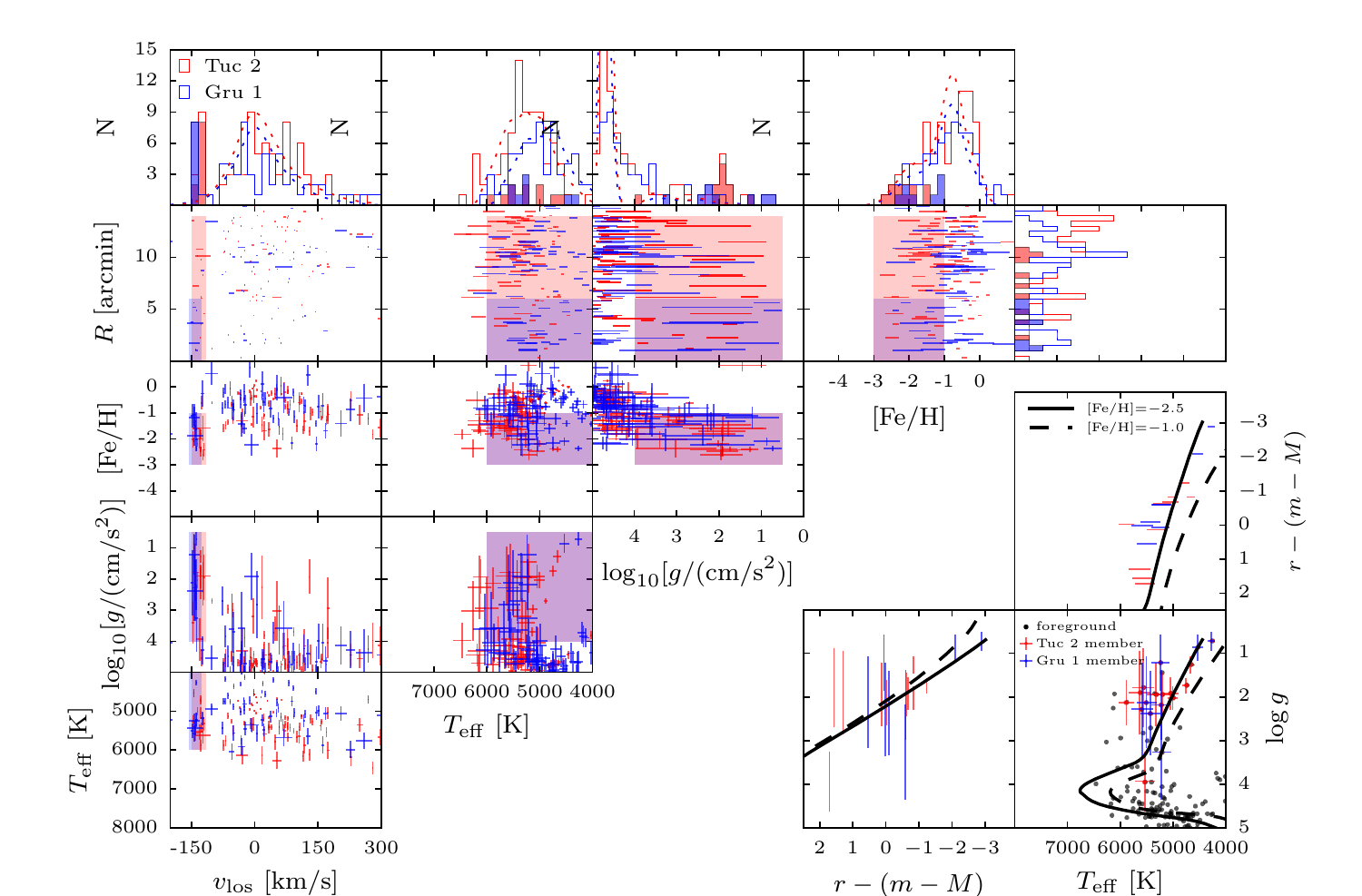}
  \caption{Scatter plots showing relations among angular separation from center ($R$) and spectroscopically-derived ($v_{\los}$, $\teff$, $\logg$ and $\feh$) quantities for individual stars along lines of sight to Tuc 2 (red) and Gru 1 (blue).  Shaded boxes enclose probable members.  For each observable, histograms show 1D distributions for the full sample (open) and likely members (filled); dotted curves indicate foreground distributions calculated from the Besan\c con Galactic model \citep{robin03}.  The three panels at bottom-right show how spectroscopic estimates of $\teff$ and $\logg$ compare with photometric magnitude and isochrone (age=12 Gyr) relations \citep{dotter08}, with red and blue markers representing probable members of Tuc 2 and Gru 1, respectively.  }
  \label{fig:tuc2gru1_params}
\end{figure*}

For all stars with formal velocity errors $\leq 20$ km s$^{-1}$, scatter plots in Figure \ref{fig:tuc2gru1_params} display relationships among spectroscopically- and photometrically-derived quantities.  

\subsection{Membership and Chemodynamical Properties}
\label{subsec:membership}

In previous analyses of similar data sets for the dwarf galaxy Reticulum 2, W15 and \citet{koposov15} employed different strategies to identify member stars and infer chemodynamical properties.  W15 applied rigid velocity, metallicity and surface gravity cuts to discard likely foreground stars, then used the remaining sample of likely members to estimate parameters of a chemodynamical model that allowed for the velocity and metallicity gradients found in many dwarf galaxies \citep{battaglia06,battaglia08,battaglia11,walker08}.  In contrast, \citet{koposov15} modeled their entire sample as a mixture of member and foreground populations, but did not allow for velocity or metallicity gradients.  Here we combine both strategies, modeling our Tuc 2 and Gru 1 samples as mixtures of member and foreground populations while simultaneously allowing for gradients.  In this analysis we employ the same quality-control criteria we have used in previous work (e.g., W15), considering only the \tucqc\ stars in Tuc 2 (\gruqc\ stars in Gru 1) for which posterior PDFs for velocity are approximately Gaussian (skewness $|S|\leq 1$, kurtosis $|K-3|\leq 1$).  

\smallskip 
We assume that a fraction $f_{\rm mem}$ of a given sample is contributed by member stars whose velocities and metallicities follow a bivariate normal distribution,
\begin{equation}
  p_{\rm mem}\bigl (v_{\los},\feh |\boldsymbol{\mu}_{\rm mem},\boldsymbol{\Sigma}_{\rm mem}\bigr )=\mathcal{N}_2(\boldsymbol{\mu}_{\rm mem},\boldsymbol{\Sigma}_{\rm mem})
\end{equation}
with mean vector
\begin{equation}
  \boldsymbol{\mu}_{\rm mem}\equiv
  \left(
    \begin{matrix}
      \langle v_{\los}\rangle_{\rm mem}+k_{v_{\los}}R\cos(\theta-\theta_{v_{\los}})\\
      \langle \feh\rangle_{\rm mem}+k_{\feh}R
    \end{matrix}
  \right )
  \label{eq:membermeanvector}
\end{equation}
that depends on position $(R,\theta)$, and covariance matrix
\begin{equation}
  \boldsymbol{\Sigma}_{\rm mem}\equiv
\left(
\begin{matrix}
\sigma^2_{v_{\los},\mathrm{mem}}+\delta^2_{v_{\los}} & 0\\
0 & \sigma^2_{\feh,\mathrm{mem}}+\delta^2_{\feh}
\end{matrix}
\right ),
\end{equation}
whose diagonal elements are broadened by observational errors $\delta_{v_{\los}}$ and $\delta_{\feh}$ (we implicitly assume that velocity and metallicity are uncorrelated).  In Equation \ref{eq:membermeanvector}, $\langle v_{\los}\rangle_{\rm mem}$ is the mean velocity at the center, $k_{v_{\los}}$ is the magnitude of maximum velocity gradient and $\theta_{v_{\los}}$ (measured from north of center and opening to the east) specifies its direction.  Similarly, $\langle \feh\rangle_{\rm mem}$ is the mean metallicity at the center and $k_{\feh}$ is the magnitude of maximum metallicity gradient, which we assume to be isotropic.  

\smallskip
We further assume that the remaining fraction $1-f_{\rm mem}$ of our sample is contributed by nonmember stars whose velocities and metallicities follow a different bivariate normal distribution, 
\begin{equation}
  p_{\rm non}\bigl (v_{\los},\feh |\boldsymbol{\mu}_{\rm non},\boldsymbol{\Sigma}_{\rm non}\bigr )=\mathcal{N}_2(\boldsymbol{\mu}_{\rm non},\boldsymbol{\Sigma}_{\rm non})
\end{equation}
with mean vector
\begin{equation}
  \boldsymbol{\mu}_{\rm non}\equiv
  \left(
    \begin{matrix}
      \langle v_{\los}\rangle_{\rm non}\\
      \langle \feh\rangle_{\rm non}
    \end{matrix}
  \right )
  \label{eq:membermeanvector2}
\end{equation}
that does not vary over the M2FS field of view (radius $R_{\rm fov}\sim 29$ arcmin), and covariance matrix
\begin{equation}
  \boldsymbol{\Sigma}_{\rm mem}\equiv
\left(
\begin{matrix}
\sigma^2_{v_{\los},\mathrm{non}}+\delta^2_{v_{\los}} & 0\\
0 & \sigma^2_{\feh,\mathrm{non}}+\delta^2_{\feh}
\end{matrix}
\right ).
\end{equation}

Finally, we assume that members and nonmembers are spatially distributed according to exponential and uniform surface brightness profiles, respectively, corresponding to probability distributions
\begin{equation}
  p_{\rm mem}(R)=\frac{R}{R_{\rm e}^2}\exp\biggl [-\frac{R}{R_{\rm e}}\biggr ].
\end{equation}
and
\begin{equation}
  p_{\rm non}(R)=\frac{2R}{R^2_{\rm fov}}.
\end{equation}

\smallskip
Under these assumptions, a data set consisting of $N$ observations, $D\equiv \{(v_{\los,i},\feh_i) \}_{i=1}^N$, has likelihood
\begin{eqnarray}
  \mathcal{L}(D\big |\vec{\theta})=\displaystyle\prod_{i=1}^N\biggl ( f_{\rm mem}p_{\rm mem}(v_{\los,i},\feh_i | \boldsymbol{\mu}_{\rm mem},\boldsymbol{\Sigma}_{\rm mem}) p_{\rm mem}(R_i)\nonumber\\
  +(1-f_{\rm mem})p_{\rm non}(v_{\los,i},\feh_i\big | \boldsymbol{\mu}_{\rm non},\boldsymbol{\Sigma}_{\rm non})p_{\rm non}(R_i)\biggr ),\hspace{0.3in}
  \label{eq:likelihood}
\end{eqnarray}
that is specified by 12 free parameters: $f_{\rm mem}$, $\langle v_{\los}\rangle_{\rm mem}$, $\sigma_{v_{\los},\mathrm{mem}}$, $k_{v_{\los}}$, $\theta_{v_{\los}}$, $\langle \feh\rangle_{\rm mem}$, $\sigma_{\feh,\mathrm{mem}}$, $k_{\feh}$, $\langle v_{\los}\rangle_{\rm non}$, $\sigma_{v_{\los},\mathrm{non}}$, $\langle \feh\rangle_{\rm non}$, $\sigma_{\feh,\mathrm{non}}$.  The parameter $R_{\rm e}$ is not free, but rather is fixed at the value estimated photometrically by K15 after adjusting for ellipticity ($R_{\rm e}=6.0$ arcmin for Tuc 2, $1.0$ arcmin for Gru 1).  

\begin{figure}
  \includegraphics[width=3.5in,trim=0.25in 0in 3.0in 1.1in,clip]{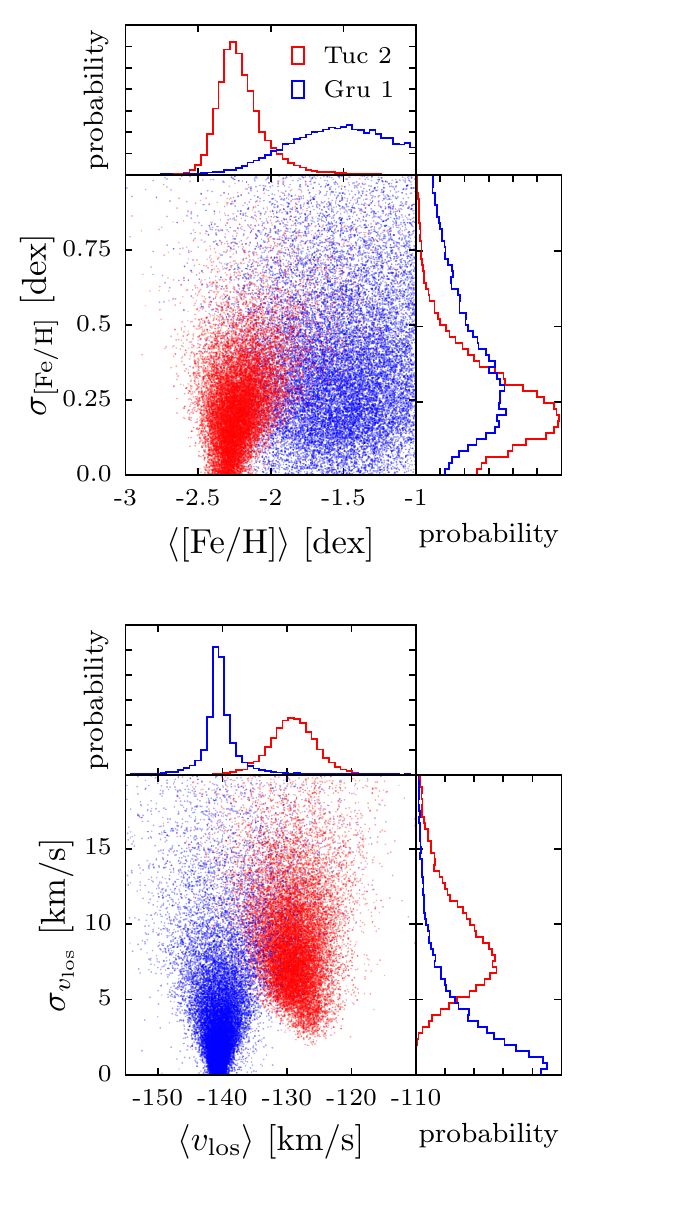}
  \caption{Samples drawn randomly from posterior PDFs for means and dispersions of metallicity (top) and velocity (bottom) distributions for Tuc 2 (red) and Gru 1 (blue).  Histograms display marginalized, 1D PDFs for each parameter.}
  \label{fig:tuc2gru1_meansdispersions}
\end{figure}

\smallskip
Table \ref{tab:tuc2gru1teffprior_mixture} lists prior PDFs that we adopt for each free parameter, as well as median-likelihood and 68\% (95\%) credible intervals of posterior PDFs (sampled by MultiNest).  Figure \ref{fig:tuc2gru1_meansdispersions} displays random samplings from posterior PDFs for the member populations' means and dispersions.
\begin{table*}
\scriptsize
\centering
\caption{Summary of probability distribution functions for chemodynamical parameters}
\begin{tabular}{@{}lllllllllll@{}}
\hline
parameter & prior & posterior: Tuc 2 & posterior: Gru 1 & description\\
\hline
\smallskip
$f_{\rm mem}$ & uniform between 0 and 1 & \tucfmemexpanded & \grufmemexpanded & member fraction\\
\smallskip
$\langle v_{\rm los}\rangle_{\rm mem}$ [km s$^{-1}$] & uniform between -160 and -110 & \tucvmeanexpanded & \gruvmeanexpanded & mean velocity at center (members)\\
\smallskip
$\sigma_{v_{\rm los},\mathrm{mem}}$ [km s$^{-1}$] & uniform between 0 and +500 & \tucvdispexpanded & \gruvdispexpanded & velocity dispersion (members)\\
\smallskip
$\langle \feh\rangle_{\rm mem} $ & uniform between -5 and -1 & \tucfehmeanexpanded & \grufehmeanexpanded & mean metallicity at center (members)\\
\smallskip
$\sigma_{\feh,\mathrm{mem}}$ & uniform between 0 and +2 & \tucfehdispexpanded & \grufehdispexpanded & metallicity dispersion (members)\\
\smallskip
$k_{v_{\rm los}}$ [km s$^{-1}$ arcmin$^{-1}$] & uniform between 0 and +10 & \tucvgradexpanded & \gruvgradexpanded & magnitude of maximum velocity gradient (members)\\
\smallskip
$\theta_{v_{\rm los}}$ [$\degr$] & uniform between -180 and +180 & \tucvthetaexpanded & \gruvthetaexpanded & direction of maximum velocity gradient (members)\\
\smallskip
$k_{\feh}$ [dex arcmin$^{-1}$] & uniform between -1 and 0 & \tucfehgradexpanded & \grufehgradexpanded & magnitude of metallicity gradient (members)\\
$\langle v_{\rm los}\rangle_{\rm non}$ [km s$^{-1}$] & uniform between -500 and +500 & \tucvmeantwoexpanded & \gruvmeantwoexpanded & mean velocity (nonmembers)\\
\smallskip
$\sigma_{v_{\rm los},\mathrm{non}}$ [km s$^{-1}$] & uniform between 0 and +500 & \tucvdisptwoexpanded & \gruvdisptwoexpanded & velocity dispersion (nonmembers)\\
\smallskip
$\langle \feh\rangle_{\rm non} $ & uniform between -5 and +1 & \tucfehmeantwoexpanded & \grufehmeantwoexpanded & mean metallicity (nonmembers)\\
\smallskip
$\sigma_{\feh,\mathrm{non}}$ & uniform between 0 and +2 & \tucfehdisptwoexpanded & \grufehdisptwoexpanded & metallicity dispersion (nonmembers)\\
\\
\hline
\end{tabular}
\\
\label{tab:tuc2gru1teffprior_mixture}
\end{table*}

\smallskip
For Tuc 2, member stars comprise a fraction \tucfmem\ of our quality-controlled ($N=$\tucqc) sample, which is sufficient to resolve a velocity dispersion of \tucvdisp\ km s$^{-1}$ about a mean of \tucvmean\ km s$^{-1}$.  The mean metallicity is \tucfehmean, but metallicity dispersion is unresolved (Figure \ref{fig:tuc2gru1_meansdispersions}).  For Gru 1, members comprise a fraction \grufmem\ of our ($N=$\gruqc) sample and we estimate a mean velocity of \gruvmean\ km s$^{-1}$, but the velocity dispersion is unresolved (Figure \ref{fig:tuc2gru1_meansdispersions}).  Gru 1 appears to be more metal-rich than Tuc 2, with a mean metallicity of \grufehmean, but the error bar is relatively large.  Both samples are too small to place meaningful limits on velocity or metallicity gradients; however, our estimates of means and dispersions marginalize over uncertainties in these gradients as well as in all other parameters (including member fraction).

\smallskip
For each individual star, we use the posterior PDFs to evaluate probability of membership, $P_{\rm member}$, which is the ratio of the first (member) term inside the product in Equation \ref{eq:likelihood} to the sum of both (member and nonmember) terms.  The last column in Table \ref{tab:tuc2gru1_table1} lists membership probability for each individual star---including those that, due to non-Gaussian velocity PDFs, were excluded from the actual likelihood calculation.  We count \tucmemtot\ stars with velocity errors $< 20$ km s$^{-1}$ that are probable members ($P_{\rm member}>0.5$) of Tuc 2, and \grumemtot\ such stars that are probable members of Gru 1.  Shaded boxes in Figure \ref{fig:tuc2gru1_params} enclose these stars, which appear as red circles (solid red if the star has a Gaussian PDF and was used in our chemodynamical analysis, otherwise open red) in the CMDs and maps of Figures \ref{fig:tuc2gru1_cmd} (note: probable members Gru1-003 and Gru1-004 are almost perfectly superimposed in color-magnitude space) and \ref{fig:tuc2gru1_map} and in the isochrone relations at the bottom-right of Figure \ref{fig:tuc2gru1_params}.  

\smallskip
We notice that five of seven probable members lie outside the projected halflight radius that K15 estimate for Gru 1 based on DES photometry.  K15 advised readers to regard this estimate with caution, given the location of Gru 1 near a chip gap in the first-year DES images.  While the configuration of probable members in Figure \ref{fig:tuc2gru1_map} suggests that Gru 1's halflight radius may be underestimated, the peculiar arrangement owes at least partially to selection: the two innermost probable members also happen to be the only stars within the halflight radius for which we obtained measurements.  Given the relatively large statistical uncertainties in our estimates of Gru 1's chemo-dynamical parameters, perhaps the only one that is usefully constrained is Gru 1's mean velocity (see Section \ref{sec:summary}).  Setting mixture models aside for the moment, if we consider only the three stars toward Gru 1 that have metallicity uncertainties $\leq 0.2$ dex (also the only three stars with median S/N $> 5$ per pixel), two have velocities within the narrow peak shaded in blue in Figure \ref{fig:tuc2gru1_params}; these same two stars are also the most metal-poor ($\feh \la -2$) in our Gru 1 sample, providing reassurance that the velocity peak we associate with Gru 1 (Figure \ref{fig:tuc2gru1_params}) is not spurious.

\subsection{Scaling Relations}
\label{subsec:scalingrelations}

\begin{figure}
  \includegraphics[width=5in,trim=0.in 0.6in 2.0in 0.5in,clip]{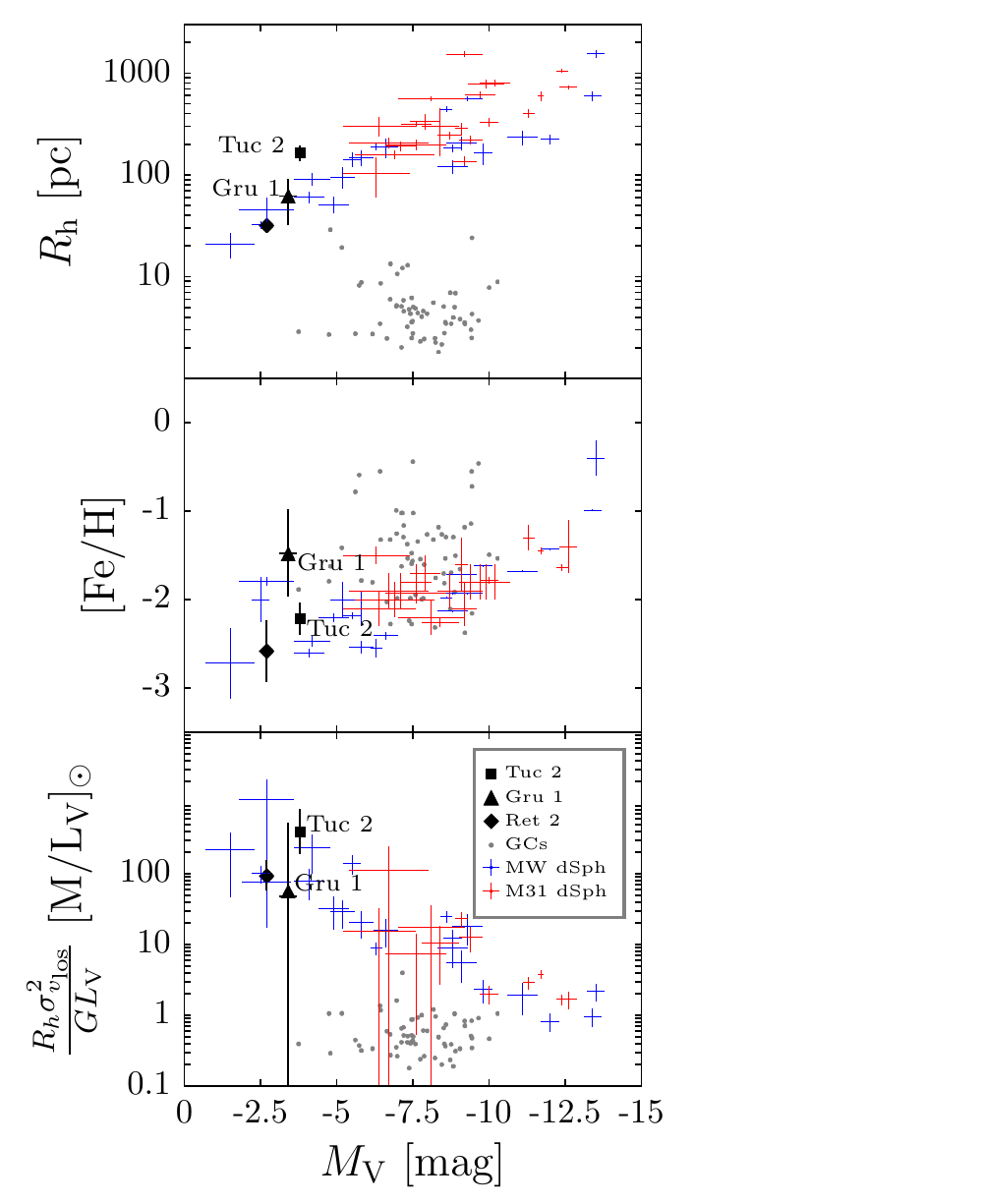}
  \caption{Size, mean metallicity and dynamical mass-to-light ratio vs absolute magnitude, for Galactic globular clusters (black points) as well as dwarf spheroidal satellites of the Milky Way (blue points with errorbars) and M31 (red points with errorbars).  Quantities plotted for Tuc 2 and Gru 1 are adopted from K15 and this work.  Data for globular clusters and dSphs are adopted, respectively, from the catalog of \citet[2010 edition; we include only clusters with velocity dispersion measurements]{harris96} and the review of \citet{mcconnachie12}.}
  \label{fig:tuc2gru1_scaling}
\end{figure}

Figure \ref{fig:tuc2gru1_scaling} places these new results in the context of scaling relations established for dwarf galaxies and globular clusters.  The mean metallicity that we measure for Tuc 2 is entirely consistent with the dwarf galaxy luminosity/metallicity (`L-Z') relation \citep[middle panel of Figure \ref{fig:tuc2gru1_scaling}]{kirby13}.  Similarly, in terms of the quantity $R_h\sigma^2_{v_{\los}}/(GL_V)$ (dimensionally a mass-to-light ratio, where $R_h$ is projected halflight radius and $L_V$ is V-band luminosity), Tuc 2 follows the well-established dwarf galaxy relation \citep{mateo98} and is separated by nearly two orders of magnitude from globular clusters at similar luminosity.  On these grounds, our results confirm that Tuc 2 is a dwarf galaxy and not a globular cluster.  

\smallskip
Our spectroscopic results are less conclusive about the nature of Gru 1.  While the mean metallicity of \grufehmean\ that we estimate for Gru 1 is typical of globular clusters, it lies within $\sim 1\sigma$ of the dwarf galaxy L-Z relation (middle panel of Figure \ref{fig:tuc2gru1_scaling}).  Furthermore, since we do not resolve Gru 1's velocity dispersion, we obtain only an upper limit on Gru 1's dynamical mass: $M(R_{\rm h})\la $ \grumrhalfupperlim$\mathrm{M}_{\odot}$ enclosed within the central $\sim 60$ pc.  This value is consistent with, but does not require, a significant dark matter component.  Pending deeper data that resolve Gru 1's velocity and metallicity dispersions, the strongest evidence that Gru 1 is a dwarf galaxy remains its large size ($R_{\rm h}\sim 60$ pc; K15), which places it on the dwarf galaxy sequence and separates it from globular clusters (top panel of Figure \ref{fig:tuc2gru1_scaling}).  

\begin{table*}
  \begin{centering}
  \scriptsize
  \caption{Summary of observed photometric and spectroscopic
    properties for Tucana 2 and Grus 1}
  \begin{tabular}{@{}lllllllllll@{}}
    \hline
    quantity&Tucana 2&Grus 1&description&reference\\
    \hline
    $\alpha_{\rm J2000}$&22:51:55&22:56:42&R.A. at center&K15$^{1}$&\\
    $\delta_{\rm J2000}$&$-$58:34:08&$-$50:09:48&Dec. at center&K15\\
    $l$ [deg]&328.0863&338.6793&Galactic longitude&K15\\
    $b$ [deg]&$-$52.3248&$-$58.2448&Galactic latitude&K15\\
    $m-M$ [mag]&$18.8\pm 0.2$&$20.4\pm 0.2$&distance modulus&K15\\
    $D$ [kpc]&$57\pm 5$&$120\pm 11$&distance from Sun&K15\\
    $M_{\rm V}$ [mag]&$-3.8\pm 0.1$&$-3.4\pm 0.3$&absolute magnitude&K15\\
    $R_{\rm e}$ [arcmin]&$7.67_{-1.18}^{+1.02}$&$1.33_{-0.26}^{+0.74}$&exponential scale length (semi-major axis)&K15\\
    $e\equiv 1-(b/a)$&$0.39^{+0.10}_{-0.20}$&$0.41_{-0.28}^{+0.20}$&ellipticity&K15\\
    PA [deg]&$107\pm 18$&$4\pm 60$&position angle&K15\\
    $R_{\rm h}$ [arcmin]&$9.83_{-1.11}^{+1.66}$&$1.77_{-0.39}^{+0.85}$&projected halflight radius$^{2}$&K15\\
    $R_{\rm h}$ [pc]&$165_{-19}^{+28}$&$62_{-14}^{+30}$&projected halflight radius$^{2}$&K15\\
\\
    $v_{\rm los}$ [km s$^{-1}$]& \tucvmean &\gruvmean & mean line-of-sight velocity, solar rest frame&this work\\
    $v_{\rm los}$ [km s$^{-1}$]&\tucvgrf & \gruvgrf & mean line-of-sight velocity, Galactic rest frame$^{3}$&this work\\
    $\sigma_{v_{\los}}$ [km s$^{-1}$]&\tucvdisp&$< $\gruvdispupperlim & internal velocity dispersion&this work\\
    $k_{v_{\los}}$ [km s$^{-1}$ arcmin$^{-1}$]&$<$ \tucvgradupperlim &$<$ \gruvgradupperlim & velocity gradient&this work\\
    $\theta_{v_{\los}}$ [deg]& \nodata &\nodata & PA of velocity gradient&this work\\
    $\langle \feh\rangle$ [dex]& \tucfehmean & \grufehmean &mean metallicity&this work\\
    $\sigma_{\feh}$ [dex]& $<$ \tucfehdispupperlim & $<$ \grufehdispupperlim &metallicity dispersion&this work\\
    $k_{\feh}$ [dex arcmin$^{-1}$]&$>$ \tucfehgradlowerlim  & $>$ \grufehgradlowerlim &metallicity gradient&this work\\
    $M(R_{\rm h})$ [$\mathrm{M}_{\odot}$]& \tucmrhalf & $<$ \grumrhalfupperlim & dynamical mass$^{4}$ enclosed within $R_{\rm h}$&this work\\
    $\Upsilon$ [$\mathrm{M}_{\odot}/L_{V,\odot}$]& \tucmlratio &$<$ \grumlratioupperlim &dynamical mass-to-light ratio$^{5}$&this work\\
    \hline
    $^{1}$ DES15 report similar values for Tuc 2.\\
    $^{2}$ $R_{\rm h}\approx 1.68 R_{\rm e}\sqrt{1-e}$\\
    $^{3}$ calculated using the solar motion measured by \citet{schonrich10}\\
    $^{4}$ $M(R_{\rm h})\approx 5R_{\rm h}\sigma_{v_{\los}}^2/(2G)$; assumes equilibrium, negligible binary stars \\
    $^{5}$ $\Upsilon\approx 2M(R_{\rm h})/L_V$\\
  \end{tabular}
  \label{tab:summarytable}
\end{centering}
\end{table*}

\subsection{Dark Matter Content of Tuc 2}
Assuming spherical symmetry, dynamic equilibrium and that binary motions contribute negligibly to the measured velocity dispersion, one can use crude approximations to translate the size and velocity dispersion of Tuc 2 into a dynamical mass.  For example, the formula of \citet{walker09d}\footnote{The formula of \citet{wolf10} is similar, implying that the mass enclosed within the 3D halflight radius, $r_{1/2}\approx 4/3 R_{\rm h}$, is $M(r_{1/2})\approx 3r_{1/2}\sigma_{v_{\los}}^2/G$.} implies that the mass enclosed within the sphere with radius equal to Tuc 2's projected halflight radius ($R_{\rm h}=165_{-18.5}^{+27.8}$ pc; K15) is $M(R_{\rm h})\approx 5R_{\rm h}\sigma_{v_{\los}}^2/(2G)=$ \tucmrhalf $\mathrm{M}_{\odot}$.  The  corresponding dynamical mass-to-light ratio is $\approx 2M(R_{\rm h})/L_V=$ \tucmlratio\ (the $99.9\%$ credibility interval excludes values smaller than $200$) in solar units, suggesting that dark matter dominates Tuc 2's gravitational potential.  

\smallskip
Finally, we use our velocity data to estimate the profiles $\deriv D/\deriv\Omega\equiv \int_0^{\infty}\rho(l)\deriv l$ and $\deriv J/\deriv\Omega\equiv \int_0^{\infty}\rho^2(l)\deriv l$, where $\rho(l)$ is the density of dark matter at distance $l$ along the line of sight.  For a given particle physics model, the quantities $D(\theta)\equiv \int_0^{\theta}2\pi\sin(\theta')\deriv\theta'\deriv D(\theta')/\deriv\Omega$ and $J(\theta)\equiv \int_0^{\theta}2\pi\sin(\theta')\deriv\theta'\deriv J(\theta')/\deriv\Omega$ are proportional to the flux of photons from dark matter decay and annihilation processes, respectively.  We estimate the density profile under assumptions of spherical symmetry and dynamic equilibrium, following the procedure described in detail by \citet{bonnivard15,bonnivard15b}.  Table \ref{tab:jfactors} lists the corresponding $D(\theta)$ and $J(\theta)$ factors and 68\% (95\%) credible intervals for a range of integration angles.  At $\theta=0.5^{\circ}$, the angle that is typical of previous gamma-ray searches using Fermi-LAT \citep{fermi14etal,geringer-sameth15c}, we obtain $\log_{10}[D(0.5^{\circ})/(\mathrm{GeV}\mathrm{cm}^{-2})]=18.4_{-0.7}^{+0.7}$ and  $\log_{10}[J(0.5^{\circ})/(\mathrm{GeV}^{2}\mathrm{cm}^{-5})]=18.7_{-0.7}^{+0.9}$.  These values rank Tuc 2 $\sim 10^{\rm th}$ among known dwarf galaxies for decay, and $\sim 8^{\rm th}$ among known dwarf galaxies for annihilation (cf. Figure 6 of \citealt{bonnivard15}, Figure 4 of \citealt{bonnivard15b}, Figure 4 of \citealt{geringer-sameth15b}).  Thus we do not expect Tuc 2 to exhibit the strongest decay/annihilation signals, but it can contribute meaningfully to analyses that stack data from multiple sources.  
\begin{table}
\begin{center}
\caption{Dark matter decay and annihilation emission factors for Tuc 2.  
\label{tab:jfactors}}
\begin{tabular}{ccccc} \hline \hline
  $\theta$ &&  $\log_{10}[D(\theta)/(\mathrm{GeV}\mathrm{cm}^{-2})]$   &&   $\log_{10}[J(\theta)/(\mathrm{GeV}^2\mathrm{cm}^{-5})]$\\[0.1cm]
 [deg] && &&\\[0.05cm]
\hline
0.01 && $15.4_{-0.3(-0.7)}^{+0.4(+0.9)}$ && $16.3_{-0.7(-1.5)}^{+0.7(+1.6)}$ \\[0.2cm]
0.05 && $16.7_{-0.4(-0.6)}^{+0.5(+1.0)}$ && $17.6_{-0.7(-1.3)}^{+0.6(+1.3)}$ \\[0.2cm]
0.1  && $17.3_{-0.4(-0.7)}^{+0.5(+1.0)}$ && $18.0_{-0.7(-1.3)}^{+0.6(+1.4)}$ \\[0.2cm]
0.5  && $18.4_{-0.7(-1.2)}^{+0.7(+1.2)}$ && $18.7_{-0.7(-1.4)}^{+0.9(+1.8)}$ \\[0.2cm]
1    && $18.8_{-0.9(-1.5)}^{+0.8(+1.4)}$ && $18.9_{-0.9(-1.5)}^{+1.1(+2.1)}$ \\[0.1cm]
\hline

\end{tabular}
\end{center}
\end{table}

\section{Summary \& Discussion}
\label{sec:summary}

Table \ref{tab:summarytable} summarizes the observed properties of Tuc 2 and Gru 1, combining our new spectroscopic results with the previous photometric results from K15 and DES15.  Our M2FS spectroscopy confirms that Tuc 2 is a dwarf galaxy, following the well-established luminosity-metallicity relation and possessing a large dynamical mass-to-light ratio.  Gru 1 is likely also to be a dwarf galaxy but, given the large uncertainty in its spectroscopic metallicity and its unresolved velocity dispersion, the strongest evidence for this conclusion remains its large size ($R_{\rm h}\sim 60$ pc; K15).  Nevertheless, our spectra are sufficient to measure Gru 1's mean velocity (\gruvmean\ km s$^{-1}$ in the solar rest frame), which is strikingly similar to that of Tuc 2 (\tucvmean\ km s$^{-1}$).  Taken with the fact that Tuc 2 and Gru 1 are separated by just $\sim 30$ kpc in three-dimensional space, these results provide new clues about the Galactic halo and/or the origins of its occupants.

\begin{figure}
    \includegraphics[width=3.in, trim=0in 0in 0.in 0in,clip]{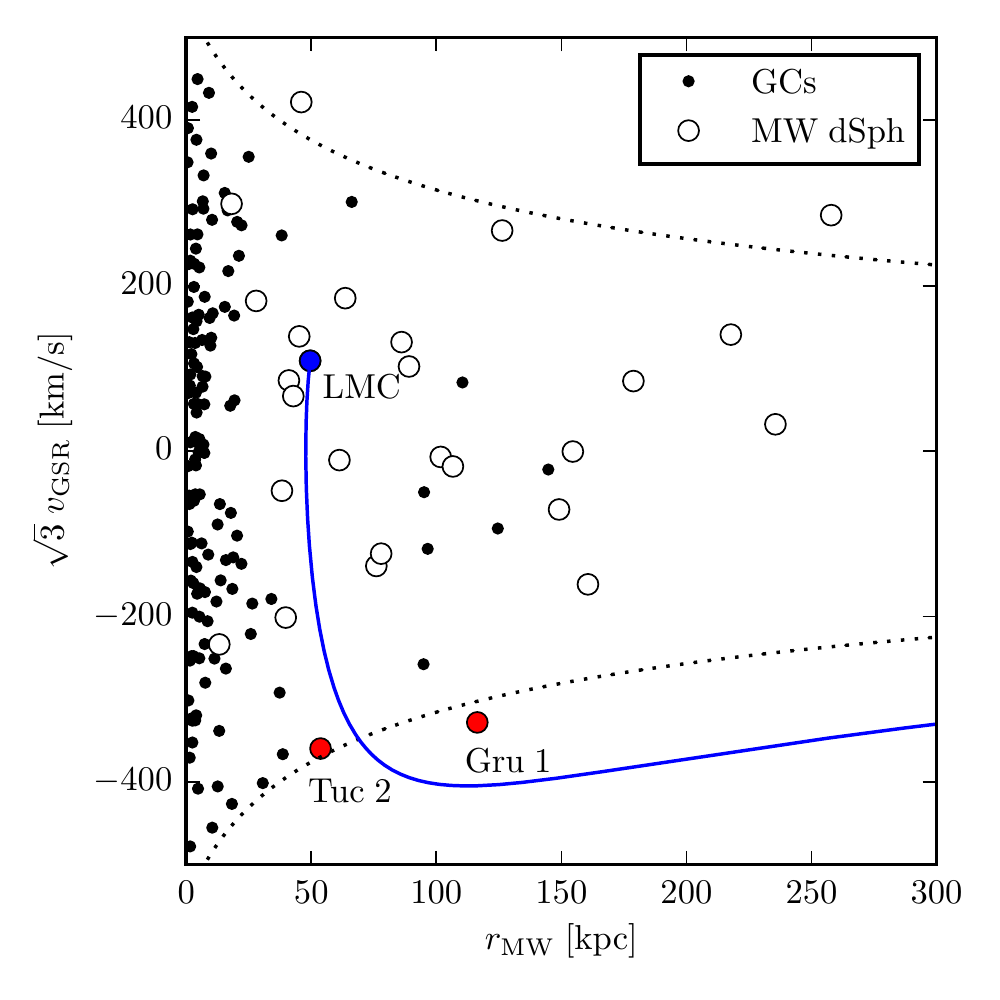}
  \caption{Line-of-sight velocity (in the Galactic standard of rest) vs Galactocentric distance, for the Milky Way's globular clusters \citep[][black dots]{harris96}), dwarf spheroidal satellites \citep[][white circles]{mcconnachie12} and Tuc 2 and Gru 1 (red circles); velocities have been multiplied by $\sqrt{3}$ to account for unknown tangential velocities. Dotted curves indicate escape velocity for a \citet{navarro97} halo with Milky-Way-like mass $M_\mathrm{200} = 10^{12} M_\odot$, and concentration $c=12$.  Solid curves represent the past orbit of the LMC, integrated in this potential.}
  \label{fig:orbits}
\end{figure}

\smallskip
For example, Figure \ref{fig:orbits} depicts coordinates in radial phase space for Galactic satellites, placing our estimates for Tuc 2 and Gru 1 among other dwarf galaxies (data from \citealt{mcconnachie12}) and globular clusters \citep[][2010 update]{harris96}.  Viewed in this space, Tuc 2 and Gru 1 appear to be among the least bound objects in the Galactic halo.  After applying a crude adjustment to account for unknown tangential velocities---i.e., multiplying by a factor of $\sqrt{3}$---the magnitude of Gru 1's velocity in the Galactic rest frame ($v_{\los}=$ \gruvgrf\ km s$^{-1}$) exceeds the local escape speed calculated for a \citet[][NFW hereafter]{navarro97} dark matter halo with Milky-Way-like mass $M_{200}=10^{12}\mathrm{M}_{\odot}$ and concentration $c=12$ \citep{klypin02}.  Only two other dwarf galaxies with distances $\la 300$ kpc are known to share this property: Leo I ($r_{\rm MW}\sim 260$ kpc, $v_{\rm GSR}\sim +170$ km s$^{-1}$; \citealt{zaritsky89}) and Bo\"otes III ($r_{\rm MW}\sim 46$ kpc, $v_{\rm GSR}\sim +240$ km s$^{-1}$; \citealt{carlin09}).  Thus, Gru 1 may join these objects in building the case for a more massive Milky Way \citep[e.g.,][]{boylan-kolchin13}.  

\smallskip
On the other hand, the relatively small phase-space separation between Tuc 2 and Gru 1 might point in a different direction.  Such a configuration would arise naturally if the two objects share similar orbital histories via a mutual association with the Magellanic Clouds.  The blue curve in Figure \ref{fig:orbits} represents the past orbit of the LMC, integrated in the potential of the aforementioned NFW halo with boundary conditions given by the proper motion measurements of \citet{kallivayalil13}.  The orbit calculation includes effects of dynamical friction within the Galactic halo, estimated using the formula of \citet{chandrasekhar43}, which we calibrate using N-body simulations (Jethwa et al., in preparation).  The radial phase-space coordinates of Tuc 2 and Gru 1 lie near what would be the trailing tail of the LMC along this orbit (Figure \ref{fig:orbits}).  

\begin{figure*}
    \includegraphics[width=6.5in, trim=0in 0in 0.in 0in,clip]{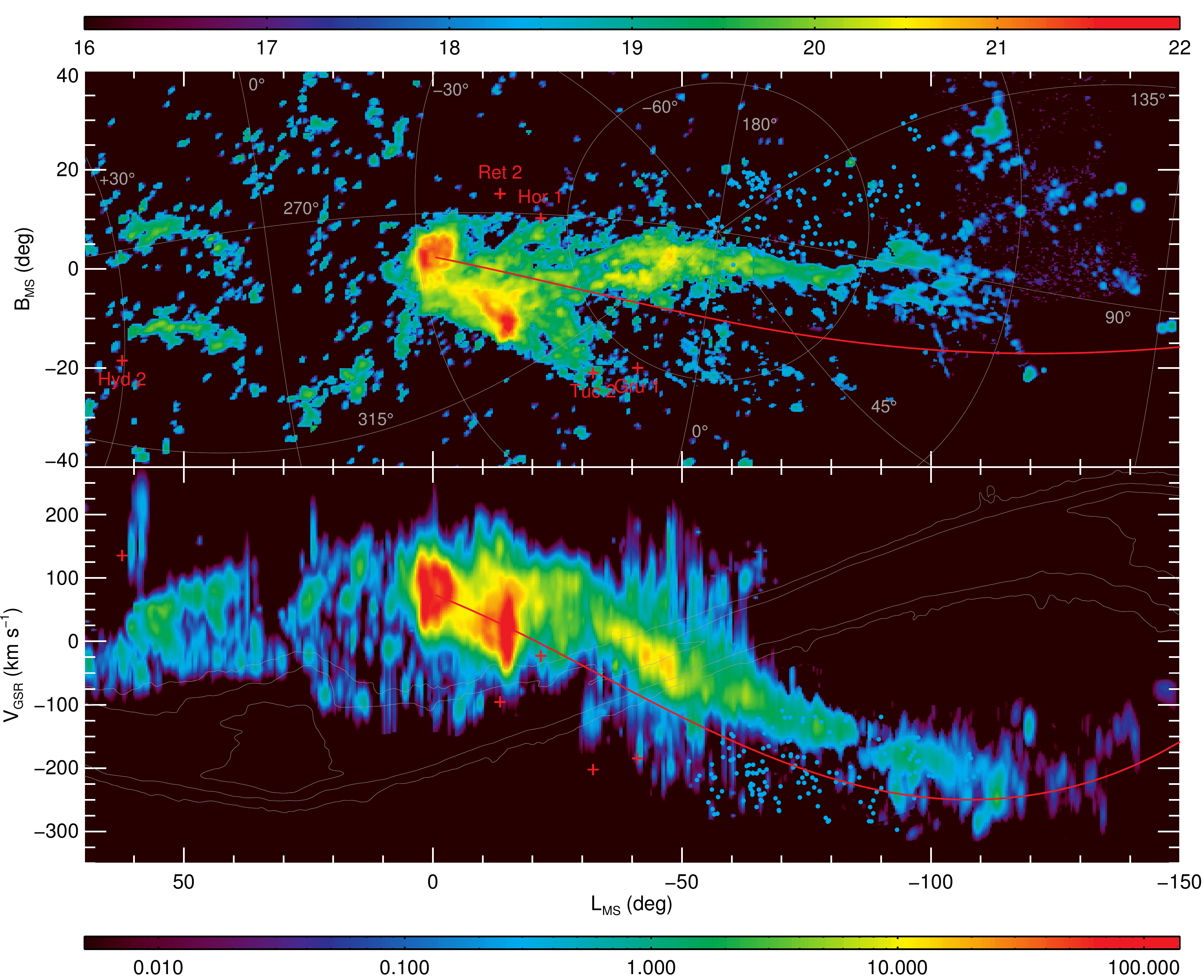}
    \caption{Positions and velocities (Galactic standard of rest) of objects near the Magellanic Stream  \citep{nidever10}.  The coordinate system has equator ($B_{\rm MS}=0$) along the stream, with longitude ($L_{\rm MS}$) decreasing along the portion trailing the LMC \citep{nidever08}.  In the top panel, color scale represents HI column density ($\log_{10}[N_{\rm HI}/(\mathrm{cm}^{-2})]$), red crosses identify newly-discovered ultrafaint objects with spectroscopic measurements, and filled blue circles are compact high velocity clouds from \citet{westmeier08}.  In the bottom panel, color scale represents total intensity of Magellanic HI integrated along $B_{\rm MS}$ (units of K), and gray contours indicate Galactic HI emission (intervals of 10 100, 1000 K).  Red curves depict the same LMC orbit shown in Figure \ref{fig:orbits}.}
  \label{fig:stream}
\end{figure*}

\smallskip
Delving more deeply into this scenario, Figure \ref{fig:stream} compares positions and velocities of Tuc 2, Gru 1 and other newly-discovered ultrafaint satellites---Horologium I \citep{koposov15betal}, Reticulum 2 \citep{koposov15betal,simon15etal,walker15b}, Hydra II \citep{kirby15}---to those of structures associated with the Magellanic Stream \citep[`MS';][]{nidever10}.  The satellite configuration straddles the Magellanic system and generally follows its velocity trend.  While Ret 2, Tuc 2 and Gru 1 move with systematically more negative velocities than does the Stream, this situation is expected for Magellanic satellites, which are impervious to the ram pressure that affects Stream gas (ram-pressure also causes trajectories of the Stream to differ from that of the LMC itself).  Moreover, positions of the satellites are offset from the MS midplane by up to $\sim 20^{\circ}$, causing their velocity vectors to project along the line of sight differently than if they were in the midplane.  Accounting for this effect, which is largest for Tuc 2 ($\sim 20$ km s$^{-1}$), would bring the satellites into closer agreement with the calculated LMC orbit.  With this correction, Gru 1's velocity deviates from that of the calculated LMC orbit (at Gru 1's stream longitude, projected along the line of sight to Gru 1) by $\sim 85$ km s$^{-1}$, and Tuc 2's deviates by $\sim 135$ km s$^{-1}$.  Both offsets are consistent with the random orbital motions expected for Magellanic satellites \citep{deason15}.  We present a thorough investigation of this scenario in forthcoming paper (Jethwa et al.\ in preparation).

\smallskip
We thank Ian Roederer, Colin Slater and Monica Valluri for helpful discussions.  We thank Jeff Crane, Steve Shectman and Ian Thompson for invaluable contributions to the design, construction and support of M2FS.  Additionally we thank an anonymous referee for helpful comments and suggestions.  MGW is supported by National Science Foundation grants AST-1313045 and AST-1412999.  MM is supported by NSF grant AST-1312997.  EWO is supported by NSF grant AST-1313006.  DLN was supported by a McLaughlin Fellowship at the University of Michigan.  The research leading to these results has received funding from the European Research Council under the European Union's Seventh Framework Programme (FP/2007-2013)/ERC Grant Agreement no. 308024.  


\end{document}